\documentclass[reprint,twocolumn,secnumarabic,amssymb,amsmath,nofootinbib,showkeys,showpacs,aps,pra,groupedaddress,superscriptaddress]{revtex4-2}
\usepackage{graphicx}

\usepackage[utf8]{inputenc}
\usepackage[english]{babel}
\usepackage{bm}
\usepackage{amsmath}
\usepackage{amsfonts}
\usepackage{amssymb}
\usepackage{epstopdf}
\usepackage{natbib}
\usepackage{rotating}
\usepackage{color}
\usepackage{dcolumn} % Align table columns on decimal point
\usepackage{ulem}
\usepackage{times}
\usepackage{booktabs}
\usepackage[colorlinks=true,linkcolor=cyan,urlcolor=cyan,citecolor=cyan]{hyperref}

\renewcommand{\vec}{\mathbf}
\newcommand{\scatlen}{a_\mathrm{S}}

\begin{document}

%\title{Quasi--bound scalar field configurations in a regular and non--regular
%black hole spacetimes}
\title{Oscillating Quantum Droplets from the free expansion of Logarithmic One--Dimensional Bose Gases}

\author{Omar Abel Rodr\'iguez-L\'opez}
\email{oarodriguez.mx@gmail.com}
\homepage{https://orcid.org/0000-0002-3635-9248}
\affiliation{Instituto de Física, Universidad Nacional Autónoma de México, Apartado Postal 20-364, 01000 Ciudad de México, México}

\author{El\'{\i}as Castellanos %\orcidlink{0000-0002-7615-5004}
}
\email{ecastellanos@mctp.mx}
\homepage{https://orcid.org/0000-0002-7615-5004}
\affiliation {Mesoamerican Centre for Theoretical Physics\\ Universidad Aut\'onoma de Chiapas.
  Ciudad Universitaria, Carretera Zapata Km. 4, Real del Bosque (Ter\'an), 29040, Tuxtla Guti\'errez, Chiapas, M\'exico.}

%--------------------------------------------------------------------------------------------------------------------------------------------
%--------------------------------------------------------------------------------------------------------------------------------------------
%\date{Modified: \today / Compiled: \today}

\begin{abstract}
  We analyze some issues related to the stability and free expansion of a one--dimensional
  logarithmic Bose--Einstein condensate, particularly its eventual relation to the formation of
  quantum droplet--type configurations.
  We prove that the corresponding properties, such as the energy of the associated N--body ground
  state, differ substantially with respect to its three--dimensional counterpart.
  Consequently, the free velocity expansion also shows differences with respect to the
  three--dimensional system when logarithmic interactions are taken into account.
  The one--dimensional logarithmic condensate tends to form quantum droplet--type configurations
  when the external trapping potential is turned off, i.e., the \textit{self--sustainability} or
  \textit{self--confinement} appears as in three--dimensions.
  However, we obtain that for some specific values of the self--interaction parameters and the
  number of particles under consideration, the cloud oscillates during the free expansion around to
  a specific equilibrium size.
  These results show that we can get scenarios in which the one--dimensional cloud reaches stable
  configurations, i.e., oscillating quantum droplets.
\end{abstract}

%\date{}
%\pacs{04.60.Bc, 04.90.+e, 05.30.Jp}

\maketitle

%--------------------------------------------------------------------------------------------------------------------------------------------
%--------------------------------------------------------------------------------------------------------------------------------------------
\section{Introduction}%
\label{sec:intro}

The emergence of quantum droplets in Bose--Einstein condensates (BECs) has stimulated many works
concerning this fascinating phenomenon~\cite{G1,G2,G3,G4,QLD}.
The formation of quantum droplets depends on the balance of attractive and repulsive interactions
within the system, which in some cases stabilize the BEC against collapse or explosion.
Additionally, it is well known that quantum fluids can show liquid or gas behavior according to the
corresponding interactions within the system, below some critical temperature.
In the case of ultracold quantum gases, several mechanisms to stabilize the system have been
proposed, for instance, quantum fluctuations and three--body correlations~\cite{QLD}.
In the case of a mixture of two BECs with competing contact interactions, ultracold atomic droplets
have also been observed~\cite{QLD,GS,PC}.
While a single--component attractive condensate with only contact interactions
collapses~\cite{cola,cola1}, quantum fluctuations stabilize a two--component mixture with
inter--component attraction and intra--component repulsion~\cite{compo}.
In the single--component scenario, quantum droplets' formation has also been analyzed, in which
three--body and higher--order interactions can be inserted as a logarithmic term in the
corresponding interacting potential.
Consequently, it can be proved that the so--called \textit{self--sustainability} or
\textit{self--confinement} (that can be interpreted as quantum droplets) appears~\cite{Self_sus}.
We must mention here that BECs with logarithmic interactions can play a central role in this
scenario since a logarithmic interaction potential can account for multi--body
interactions~\cite{Self_sus,BB,KG}; consequently, they can use to describe dense systems.
However, the results mentioned above have been obtained for three--dimensional systems. We have to
mention here that the dynamics of one--dimensional quantum droplets have been analyzed, for
instance, in Ref.~\cite{GEA}.
However, in the latter work, the authors study the behavior of one--dimensional binary Bose gases,
nor a single component with a specific interacting potential, i.e., a logarithmic interaction
potential that can encode the contributions of multi--body interactions in a single sample.

Let us remark that BECs in one or two spatial dimensions open up a very interesting scenario to
analyze the free expansion of the cloud and its relationship with the search of quantum--droplet
type configurations.
It is well known that the physics related to low--dimensional BECs contains essential departures
from its three--dimensional counterpart.
Although (strictly speaking) one--dimensional BECs can never be achieved, it is possible to obtain
quasi--one--dimensional BECs in the laboratory using extremely anisotropic traps.
For instance, Refs.~\cite{1D,1D1,1D2,1D3} fulfill the theoretical predictions in the full
one--dimensional theoretical description under certain conditions.
Also, it must be mentioned that one--dimensional BECs have some pathological behavior in the
thermodynamic limit~\cite{PATO,PATO1,PATO2,mio}.
Finite--size effects in the system are required to make the condensation possible, and consequently,
the ground state energy per particle contributions must be taken into account.
Additionally, the study of one--dimensional BECs shows that these systems could lie in the
high--density regime~\cite{mio,601,602} and suggests some Bose--Fermi duality in one--dimensional
systems~\cite{dual,dual1,dual2,dual3,dual4}.
Thus, according to this point of view, three--body interactions (and clearly, higher--order
interactions) can be relevant in the stability analysis of the one--dimensional cloud and could also
be relevant in the eventual formation of quantum droplets in the one--dimensional regime for a
single component.

Some properties associated with BECs, particularly those associated with their stability, can be
significantly modified by the interatomic interactions.
For instance, it is possible to tune the scattering length $\scatlen$ (which describes the
two--particle interactions in the mean--field approach) using Feshbach's resonances.
This way, the interactions can be tuned from the repulsive regime $(\scatlen>0)$ through the ideal
gas $(\scatlen \sim 0)$ and finally to the attractive regime $(\scatlen < 0)$, where the gas becomes
unstable and collapses~\cite{JL,cola1}.
Thus, a logarithmic interaction potential, which generalizes the system's interactions, must be
relevant in the stability analysis of one--dimensional BECs and a fundamental ingredient in quantum
droplets' formation.
Additionally, it is noteworthy to mention that one of the more interesting phenomena related to BECs
is the free expansion of the condensate and the emergence of interference fringes of two overlapping
BECs when the trapping potential is turned off~\cite{MR}.
Let us remark that when the trapping potential is turned off, the free velocity expansion of the
three--dimensional BEC and also for the one--dimensional counterpart corresponds approximately to
the velocity predicted by Heisenberg's uncertainty principle in the ideal case, i.e.,
$\scatlen=0$~\cite{MR,Pethick}.
However, when interactions are present, this situation could be drastically affected, leading to the
system's collapse or implosion after the expansion under certain circumstances~\cite{cola1}.
Moreover, as we describe in our work, interactions can stabilize the system, so
\textit{self–sustainability} or \textit{self–confinement} can be reached. Therefore, the system can
form stable configurations or quantum droplet--type configurations, even when the trapping potential
is turned off.

In the present work, we investigate the stability conditions for a one--dimensional BEC when
logarithmic interactions within the system are present.
We analyze how the insertion of this type of logarithmic self--interaction potential modifies the
dynamics of the free expansion of the cloud and its relationship with the eventual formation of
oscillating quantum droplet--type configurations.
This paper is organized as follows: In Section~\ref{sec:1DBEC}, we present some properties related
to one dimensional systems.
In Section~\ref{sec:stability-conditions}, we analyze the stability conditions associated with
one--dimensional BECs, and consequently, we explore the free expansion of the system when
logarithmic self--interactions are present.
In Section~\ref{sec:free-expansion}, we analyze the relation between free expansion and the
formation of quantum droplet--type configurations.
Finally, in Section~\ref{sec:5}, we present a discussion and the main results.
%

%%% Added by oarodriguez
\section{One--dimensional Logarithmic Bose--Einstein Condensate}%
\label{sec:1DBEC}

The time-dependent behavior of BECs gives relevant information concerning its dynamics, regarding,
for instance, the free expansion of the corresponding cloud when the trapping potential is turning
off.
The Gross--Pitaevskii equation that describes the BEC's dynamics, under certain considerations, can
be obtained formally by using a variational formulation.
For instance, in three--dimensions, the Gross--Pitaevskii equation can be obtained through the action
principle
\begin{equation}
  \label{eq:L}
  \delta\int _{t_{1}}^{t_{2}}Ldt=0,
\end{equation}
where $L$ is the corresponding Lagrangian, given by
\begin{equation}
  \label{eq:L!}
  L=\int d\vec r \frac{i \hbar}{2}\Bigl(\Psi^{*}\frac{\partial \Psi}{\partial t}-\Psi \frac{\partial \Psi^{*}}{\partial t}\Bigr)-E,
\end{equation}
%
% Copied from E. Castellanos
where $\Psi$ is the BEC wave function or the so--called order parameter in the three--dimensional
scenario. For our purposes, the energy $E$ including the contribution of the logarithmic potential
in the above equation is defined as
\begin{align}
  \nonumber E & = \int d\vec r \left\lbrace \frac{\hbar^2}{2m} {\left| \nabla \Psi \right|}^2 + V(\vec r) {\left| \Psi \right|}^2 + \right.
  \\
  \label{eq:log-erg-func-3d}
              & \quad + \left. \frac{1}{2} g_{\mathrm{3D}} {\left| \Psi \right|}^4 - \beta  |\Psi|^2 \left[ \ln\left(\alpha^3 {\left| \Psi  \right|}^2\right) - 1 \right] \right\rbrace,
\end{align}
i.e., the dilute BEC properties in three--dimensions with logarithmic interactions are governed by
the later energy functional as in Ref.~\cite{Claus}.

Notice that in the functional form of the logarithmic potential, namely
\begin{equation}
  \label{eq:LOGPOT0}
  V_{\beta}=- \beta  |\Psi|^2 \Bigl[ \ln \Bigl(\alpha^3 {\left| \Psi  \right|}^2\Bigr) - 1\Bigr],
\end{equation}
we can identify $|\Psi|^2$ as the density of particles.
Moreover, it must be mentioned that the logarithmic potential Eq.~\eqref{eq:LOGPOT0} is regular at
$n=0$, as was mentioned in Ref.~\cite{Self_sus}, and always has the Mexican--hat shape as a function
of the order parameter.
Additionally, it is straightforward to show that the logarithmic potential Eq.~\eqref{eq:LOGPOT0}
admits a power expansion around $1/\alpha^{3}$, i.e., a perturbative limit that describes several
orders in the self--interactions.
%In this perturbative limit it is possible to obtain the so--called Gross--Pitaevskii--Ginzburg
%equation as the which describes three--body interactions, see for instance Ref.~\cite{Xi} and
%references therein.
However, the expansion mentioned above is not the general case. Thus, the logarithmic nonlinearity
can be interpreted as a potential for describing the multi--body interactions within the system, and
the analysis must be, in general, non--perturbative.

Consequently, after the variation of the energy functional Eq.~\eqref{eq:log-erg-func-3d}, the
corresponding logarithmic Gross--Pitaevskii equation (LogGPE) can be obtained as
\begin{align}
  \nonumber i \hbar \frac{\partial \Psi(\vec r, t)}{\partial t} & = \left[ - \frac{\hbar^2}{2m} \nabla^2 + V(\vec r) g_{\mathrm{3D}} {\left| \Psi(\vec r, t) \right|}^2 \right. \; -
  \\
    \label{eq:log-gpe-3d}
                                                                & \quad - \left. \beta \ln\left(\alpha^3 {\left| \Psi(\vec r, t) \right|}^2\right) \right]  \Psi(\vec r, t),
\end{align}
subject to the following condition
\begin{equation}
  \int {\left| \Psi(\vec r,t) \right|}^2 \; d\vec r = N.
\end{equation}
In the above equations, $m$ is the mass of the corresponding boson, $V(\vec r)$ is the external
potential, $g_{\mathrm{3D}} = 4 \pi \hbar^2 \scatlen / m$ is the interaction strength between any
pair of bosons in three--dimensions, with $\scatlen$ being the s--wave scattering length  of the
corresponding gas, $N$ is the number of particles, while $\beta$ and $\alpha^3$ measure the strength
of the nonlinear logarithmic interaction.
We must mention at this point that the LogGPE preserves all the properties associated with
density--dependent nonlinearities, which is, in fact, the case of the logarithmic potential, such as
conservation of probability and invariance under permutation.

We consider here that the three--dimensional BEC is confined in a trap described by a harmonic
potential of the form $V(\vec r) = m (\omega_x^2 x^2 + \omega_y^2 y^2 + \omega_z^2 z^2) / 2$, where
$\omega_{x,y,x}$ are the corresponding trap frequencies.
One--dimensional BECs have very elongated, cylindrical geometries, i.e., they have a cigar--like
shape.
These geometries are produced using harmonic traps with one frequency in the axial direction,
$\omega_z$, and other in the radial (transverse) direction, $\omega_{\bot} = \omega_x = \omega_y$,
such that $\omega_z \ll \omega_\bot$, i.e., very anisotropic traps.
Under these conditions, it is possible to freeze the motion of the bosons in the radial direction,
so they occupy only the ground state of the harmonic oscillator in the transverse direction.
However, the interaction energy should not excite other radial modes, which requires that
$g_\mathrm{3D} n_0 \ll \hbar \omega_\bot$ and $\beta \ll \hbar \omega_\bot$, where $n_0$ is the
condensate's peak density.
Then, the BEC wave function $\Psi(\vec r,t)$ can be factorized as
\begin{align}
  \label{eq:cigar-trap-ansatz}
  \Psi(\vec r, t)   & = e^{-i \hbar \omega_\bot t} \psi_{\bot}(x, y) \psi(z, t),     \\
  \psi_{\bot}(x, y) & = \frac{1}{\sqrt{\pi} a_\bot} e^{- (x^2 + y^2) / 2 a_\bot^2 },
\end{align}
where the radial wave function is appropriately normalized, $\int d \vec{r}_\bot {\left|
    \psi_{\bot}(x, y) \right|}^2 = 1$, and $a_\bot = \sqrt{\hbar / m\omega_\bot}$ is the
characteristic length of the trap in the radial direction.
We substitute Eq.~\eqref{eq:cigar-trap-ansatz} in Eq.~\eqref{eq:log-erg-func-3d} and then integrate
over the transverse direction, so the energy functional becomes $E = \hbar \omega_{\bot} +
  E_{\mathrm{1D}}$, where
\begin{align}
  \nonumber E_{\mathrm{1D}} & = \int dz\, \left\lbrace \frac{\hbar^2}{2m} {\left| \frac{d\psi(z, t)}{dz} \right|}^2 + V(z) {\left| \psi(z, t) \right|}^2 + \right.
  \\
  \nonumber
                            & + \frac{1}{2} g {\left| \psi(z, t) \right|}^4                                                                                                     \\
  \label{eq:log-erg-func-1d}
                            & - \beta \left. |\psi(z, t)|^2 \left[ \ln\left(\frac{\alpha^3}{N \pi e a_{\bot}^2} {\left| \psi(z, t) \right|}^2\right) - 1 \right] \right\rbrace,
\end{align}
is the one--dimensional energy. Then, the one--dimensional LogGPE and the associated normalization
condition can be expressed as
\begin{align}
  \nonumber & i \hbar \frac{\partial \psi(z, t)}{\partial t} = \left[ - \frac{\hbar^2}{2m} \frac{d^2}{dz^2} + V(z) + g {\left| \psi(z, t) \right|}^2 \right. \; -
  \\
    \label{eq:log-gpe-1d}
            & \quad \left. - \; \beta \ln\left(\frac{\alpha^3}{N \pi e a_{\bot}^2} {\left| \psi(z, t) \right|}^2 \right) \right] \psi(z, t),
  \\
            & \int {\left| \psi(z, t) \right|}^2 \; dz = N.
\end{align}
The strong trapping in the transverse direction modifies the interaction factor between bosons in
one--dimension, which becomes
\begin{equation}
  g = \frac{2 \hbar^2 \scatlen}{m a_\bot^2} = 2 \hbar \omega_\bot a_{S}.
\end{equation}
Remarkably, the integration in the transverse directions (or the dimensional reduction) reveals that
the trapping does not modify the factor $\beta$, i.e., its value in one--dimension and
three--dimensions remains the same unlike what occurs to the $g_{3D}$ parameter.
On the other hand, the integration introduces several pre--factors upon the density in the
logarithmic term.
These terms can be re--absorbed in a new effective parameter $\alpha \to {\alpha^3}/{N \pi e
  a_{\bot}^2}$ for one--dimensional clouds without loss of generalization, i.e., they can be
interpreted as an energy shift and can be re--absorbed in the total energy.
In other words, we recover the three--dimensional behavior for $\alpha^{3}$ and $\beta$ reported in
Ref.~\cite{Claus}.
Consequently, the upper bound for $\beta \leq 3.3 \times 10^{-15}$ eV obtained in
Ref.~\cite{Gahler1981} can be taken as a good approximation for our model, and $\alpha$ %apparently
is also not relevant in the one--dimensional context.

Finally, notice that the time--independent version of the one--dimensional LogGPE can be
obtained by using the stationary condition $\psi(z,t)=\psi(z)\exp{(-i \mu t/\hbar)}$
\begin{align}
  \nonumber
  \biggl[ -\frac{\hbar^{2}}{2m}\frac{d}{dz} & + V(z) + g|\psi(z)|^{2} \; -                                    \\
  \label{eq:GP}
                                            & -\beta \ln(\alpha |\psi(z)|^{2}) \biggr]\psi(z)  = \mu \psi(z),
\end{align}
which can also be derived from the time--independent one--dimensional energy functional, namely
\begin{eqnarray}
  \nonumber
  E(\psi)=\int d z &\Biggl[&\frac{\hbar^{2}}{2m} \Bigl|\frac{d\psi(z)}{dx} \Bigr|^{2}+V(z)|\psi(z)|^{2} \, + \\
  \label{eq:EN0}
  &+&U(|\psi(z)|^{2})\Biggr],
\end{eqnarray}
where the logarithmic potential in one--dimension is given by
\begin{align}
  \nonumber
  U(|\psi(z)|^{2}) & =\frac{1}{2}g|\psi(z)|^{4} \;-                         \\
  \label{eq:LOG1}
                   & -\beta |\psi(z)|^{2}\,[ \ln (\alpha |\psi(z)|^{2})-1],
\end{align}
together with $|\psi(z)|^{2}$ the corresponding density of particles and the trapping potential
described as $V(z)=m\omega_z^{2}z^{2}/2$.

%
% \section{Free expansion of the logarithmic one--dimensional condensate}
\section{Stability Conditions}%
\label{sec:stability-conditions}

We must mention here that to calculate the corresponding energy in Eq.~\eqref{eq:EN0} formally, we
have to solve the corresponding one--dimensional LogGPE Eq.~\eqref{eq:log-gpe-1d}.
However, in order to simplify the calculations, we are capable of employing an accurate expression
for the total energy of the cloud that can be obtained by using, as usual, an \textit{ansatz} of the
form~\cite{Pethick}
\begin{equation}
  \label{eq:TF}
  \psi(z)=\frac{ N^{1/2}}{{\pi}^{1/4}\sqrt{l}} \exp(-z^{2}/2\,l^{2})\exp(i\phi(z)).
\end{equation}
This ansatz is a solution of the Schr\"odinger equation associated with the non--interacting system
in one dimension, together with $N$ the corresponding number of particles.
Additionally, we interpreted the characteristic length $l = \sqrt{\hbar/m \omega_{z}}$ as the
initial size of the quasi--one--dimensional condensate in the non--interacting case.
The choice of the \textit{ansatz} Eq.~\eqref{eq:TF}, for the case of a one--dimensional LogBEC
trapped in a one--dimensional harmonic oscillator potential seems to be a reasonable conjecture.
In other words, it is clear that the \textit{ansatz} Eq.~\eqref{eq:TF} reflects the symmetry of the
trap and, in the non--interacting case, is the exact solution of the corresponding equation of
motion.
Thus, the free velocity expansion can be calculated in this scenario without loss of generality,
using the aforementioned \textit{ansatz} at least to first--order approximation to obtain the
contributions caused by the parameters related to the logarithmic interacting potential.
Let us mention at this point that the system's analysis by solving directly the corresponding
one--dimensional LogGPE deserves a more in--depth study that we will present elsewhere.

As was mentioned above, the \textit{ansatz} Eq.~\eqref{eq:TF} corresponds to the Schr\"odinger
equation's solution associated with non--interacting systems, where the phase $\phi$ can be
associated with particle currents~\cite{Pethick}.
Thus, by inserting the \textit{ansatz} Eq.~\eqref{eq:TF} in the energy functional
Eq.~\eqref{eq:EN0}, we can obtain the corresponding energy
\begin{equation}
  \label{TE}
  E=E_{F}+E_{R},
\end{equation}
where $E_{F}$ is the kinetic energy associated with particle currents
\begin{equation}
  \label{eq:EFL}
  E_{F}=\frac{\hbar^{2}}{2m} \int dz \, |\psi(z)|^{2} {\Biggl(\frac{d}{dz} \phi(z)\Biggr)}^{2}.
\end{equation}
Additionally, $E_{R}$ can be interpreted as the energy associated with an effective potential equal
to the condensate's total energy when the phase $\phi$ does not vary in space.
The term $E_{R}$ contains the contributions of the zero-point energy ($E_{0}$), the harmonic
oscillator potential ($E_{P}$), and the contributions due to the interactions among the particles
within the condensate $(E_{I})$, i.e.,
\begin{equation}
  E_{R}=E_{0}+E_{P}+E_{I},
\end{equation}
where
\begin{equation}
  \label{EZP}
  E_{0}=\frac{\hbar^{2}}{2m}\int dz \,\Bigl|\frac{d\psi(z)}{dx} \Bigr|^{2},
\end{equation}
\begin{equation}
  E_{P}=\frac{1}{2}m\omega_z^{2}\int dz\, z^{2} \, |\psi (z)| ^{2},
\end{equation}
and
\begin{eqnarray}
  \nonumber
  E_{I}&=&\frac{1}{2}g\int dz \, |\psi (z)| ^{4} \\
  \label{EA}
  &-&\beta \int dz \, |\psi (z)| ^{2}\Bigl[ \ln (\alpha \, |\psi (z)| ^{2})-1\Bigr].
\end{eqnarray}
Consequently, $E_{R}$ can be written as
\begin{eqnarray}
  \label{eq:eff-pot-erg}
  E_{R}&=&\frac{{\hbar}^{2}N}{8m l^{2}}+\frac{m{{\omega_z}}^{2} l^{2}N}{8}
  +\frac{g\,N^{2}}{4\sqrt{2\pi}\, l}\\ \nonumber &-& \frac{\beta}{4}  \left(2 \ln \left(\frac{\alpha N}{l}\right)+\sqrt{2}-2-\ln (\pi )\right)N,
\end{eqnarray}
where we have used the trial function Eq.~\eqref{eq:TF} together with Eqs.~\eqref{EZP}--\eqref{EA}
in order to obtain the above expression.
The system's equilibrium radius $l_{0}$ can be calculated by minimizing the energy $E_{R}$ in
Eq.~\eqref{TE}, that is, ${\left({d E_R}/{dl}\right)}_{l = l_0} = 0$.
Additionally, the kinetic energy contribution Eq.~\eqref{eq:EFL} is positive definite and is zero
when the phase $\phi$ is constant~\cite{Pethick}.

Since the BEC's total energy is $E = \hbar \omega_\bot + E_{\mathrm{1D}}$, it is clear that
$\hbar \omega_\bot$ arises as characteristic energy that we can use to rewrite the energy $E_R$ in
a dimensionless form.
After some algebraic steps, we can express the energy per boson as
\begin{align}
  \nonumber
  \frac{E_R}{N \hbar \omega_\bot} & = \frac{r^{-2}}{8} + \frac{\lambda^2 r^2}{8} + \frac{1}{2\sqrt{2\pi}} \left( N \frac{a_S}{a_\bot}\right) r^{-1} - \\
  \label{eq:eff-pot-erg-nodim}
                                  & \quad- \frac{\beta}{4 \hbar \omega_\bot} \left[ -2 \ln(r)  + \sqrt{2} - 2 - \ln(\pi) \right],
\end{align}
where the ratio $r = l / a_\bot$ measures the {BEC}'s size and $\lambda = \omega_z / \omega_\bot$.
Since the constant $\alpha$ is positive but arbitrary and has no physical relevance~\cite{Claus}, we
choose $\alpha = a_\bot / N$, so the contributions of the term $\ln(\alpha N / a_\bot)$ that arises
in Eq.~\eqref{eq:eff-pot-erg} can be neglected for all practical purposes. %(¿ESTO ESTÁ BIEN?)

%\textbf{Diferencias entre fisica de 3D y de 1D...Finite size corrections and stability.... explain... }

Let us point out some substantial differences between the energy Eq.~\eqref{eq:eff-pot-erg-nodim}
compared with its counterpart in three--dimensions.
The first significative difference relies on the functional form of the third right--hand term of
Eq.~\eqref{eq:eff-pot-erg}, that is, the contribution caused by the two--body interactions when
$\beta$ is set to be zero.
Due to the system's dimensionality, the above third right--hand term in
Eq.~\eqref{eq:eff-pot-erg-nodim} behaves as $r^{-1}$ instead of the usual $r^{-3}$ behavior for the
three--dimensional scenario~\cite{Pethick}.
This difference is significant when we consider a BEC with attractive interactions (such as
Lithium).
In the three--dimensional case, the interaction energy becomes the dominant term when $r \to 0$,
since its absolute value eventually becomes greater than the zero--point energy term, which behaves
as $r^{-2}$.
Then, for attractive interactions, the energy diverges to minus infinity, which is why the BEC
collapses when the number of particles is large enough.
However, in the one--dimensional scheme, the behavior is entirely different since the zero--point
energy term becomes dominant as $r \to 0$.
Even with attractive interactions, the energy reaches a global minimum and eventually becomes
positive and diverges as $r$ gets smaller.
Nevertheless, the energy can become negative if the number of bosons is large enough, a sign of
instability similar to what occurs in a three--dimensional BEC.\

\begin{figure}[t!]
  \centering
  \includegraphics[]{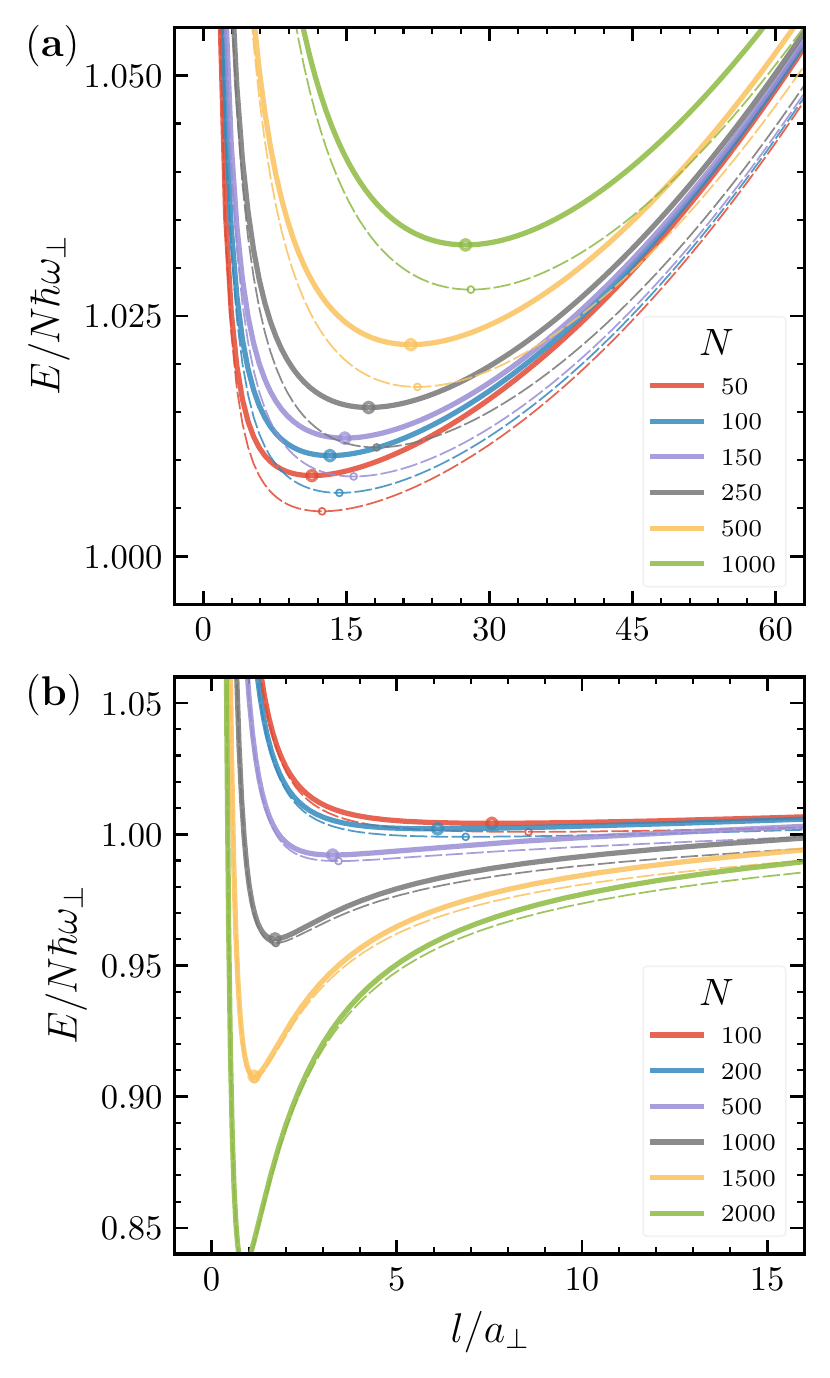}
  \caption{Energy of a logarithmic one--dimensional BEC as a function of $l / a_\bot$, for several
    numbers of particles $N$. {(a)} The scattering length is $\scatlen = 53.65 a_0$, the same as $^{23}$Na,
    see Ref.~\cite{Gorlitz2001}. (b) The scattering length is $\scatlen = -27.3 a_0$, the same as
    $^{7}$Li, see Ref.~\cite{Bradley1995}. The minimum of $E$ determines the equilibrium position
    $l_0$. In both figures, the harmonic trap frequencies are  $\omega_z / 2\pi = 3.5$ Hz and
    $\omega_\bot / 2\pi = 360$ Hz, while $\beta = 2.2174 \times 10^{-3} (\hbar \omega_\bot)$. The
    thin dashed lines show the energy of the BEC without the nonlinear interaction.}%
  \label{fig:erg_sodium-lithium_fig-01}
\end{figure}

On the one hand, in Fig.~\ref{fig:erg_sodium-lithium_fig-01}a, we show the energy of a one--dimensional
LogBEC with the characteristics reported in Ref.~\cite{Gorlitz2001} for an experiment with a gas of
$^{23}$Na.
The scattering length of sodium is $\scatlen = 53.65 a_0 = 2.80$ nm, being $a_0$ the Bohr radius,
and the BEC is confined in a trap with $\omega_z / 2\pi = 3.5$ Hz and $\omega_\bot / 2\pi = 360$ Hz.
Accordingly, we find that $a_\bot \sim 1.105$ $\mu$m, $a_z \sim 10.142 a_\bot \sim 11.207$ $\mu$m,
and $a_S / a_\bot \sim 2.534 \times 10^{-3}$.
On the other hand, optical diffraction experiments with neutrons~\cite{Gahler1981} set an upper
bound for the magnitude of the logarithmic nonlinearity in three--dimensions, which turns out to be
$\beta \leq 3.3 \times 10^{-15}$ eV, i.e., $\beta \leq 2.2174 \times 10^{-3} (\hbar \omega_\bot)$.
The results obtained in Fig.~\ref{fig:erg_sodium-lithium_fig-01}a show how the number of particles
in the condensate affects the corresponding energy landscape in the case of $^{23}$Na.
We see that the one--dimensional LogBEC energy grows as the number of particles in the gas
increases.
The minimum energy corresponding to the equilibrium size $l_0$ shows that the cloud's extent also
increases as $N$ grows.
We also can see that the logarithmic nonlinearity in the LogGPE increases the energy.
Due to this, the cloud's equilibrium size (filled circles) is slightly smaller than the equilibrium
size when the logarithmic term is turned off, i.e., $\beta = 0$ (empty circles).
All of these results are as expected since the interactions between the sodium particles are
repulsive.

On the other hand, in Fig.~\ref{fig:erg_sodium-lithium_fig-01}b, we show the corresponding energy of
a $^{7}$Li one--dimensional LogBEC, subject to a harmonic trap with the same characteristics used
for the $^{23}$Na BEC in Fig.~\ref{fig:erg_sodium-lithium_fig-01}a.
According to Ref.~\cite{Bradley1995}, Lithium's scattering length is $\scatlen = -27.3 a_0 \approx
  1.444$ nm, so $|\scatlen| / a_\bot \approx 7.22 \times 10^{-4}$.
It is clear how the attractive interactions lower the energy more and more as the number of bosons
increases since the interaction energy is negative.
It is expected that the energy becomes negative if $N$ becomes large enough, which would signal the
one--dimensional LogBEC collapse.
However, this situation can not be reliably studied using the one--dimensional LogGPE since the
\textit{ansatz} Eq.\,\eqref{eq:cigar-trap-ansatz} used for the dimensional reduction procedure
requires that $g_{\mathrm{3D}} n_0 \ll \hbar \omega_\bot$, with $n_0$ being the peak density at the
trap center, a condition that is equivalent to
\begin{equation}
  \label{eq:mean-field-cond}
  N \frac{ |\scatlen| }{l_0} = \left( N \frac{ |\scatlen| }{a_\bot} \right) {\left(\frac{l_0}{a_\bot}\right)}^{-1} \ll 1.
\end{equation}
For a large number of bosons, this condition breaks down as the interaction energy becomes
significant and comparable to $\hbar \omega_\bot$.
Hence, the bosons begin to populate the excited states in the radial direction, so the
\textit{ansatz} Eq.~\eqref{eq:cigar-trap-ansatz} is not longer appropriated.
In other words, the procedure to obtain the one--dimensional LogGPE is no longer valid.
For comparison purposes, in Fig.~\ref{fig:erg_sodium-lithium_fig-01}b, every energy curve fulfills
Eq.~\eqref{eq:mean-field-cond} at its minimum, i.e., at the one--dimensional LogBEC equilibrium
size, except for $N = 1500$ and $N=2000$, where the condition Eq.~\eqref{eq:mean-field-cond} breaks.
Therefore, the one--dimensional LogGPE energy functional Eq.~\eqref{eq:log-erg-func-1d} becomes
unsuitable for such a large number of bosons.

It must be mentioned that the above results are consistent with the behavior of low dimensional
BECs, in the sense that finite--size corrections upon the system must be taken into account to get a
well--defined condensate at finite temperature.
In other words, the system becomes unstable for a large number of particles, and it seems to be that
finite size corrections upon the one--dimensional LogBEC are necessary to get stability.
Consequently, these finite--size corrections play a central role in the formation and stability of
quantum droplet--type configurations.
%
%\textbf{This is consistent with finite-size corrections...}
%
We can conclude that the condition Eq.~\eqref{eq:mean-field-cond} guarantees that a one--dimensional
LogBEC in the regime discussed in the preceding paragraphs is stable and does not collapse even in
the presence of attractive interactions for finite--size systems.

\section{Free expansion and quantum droplets}%
\label{sec:free-expansion}

\begin{figure}[b!]
  \centering
  \includegraphics[]{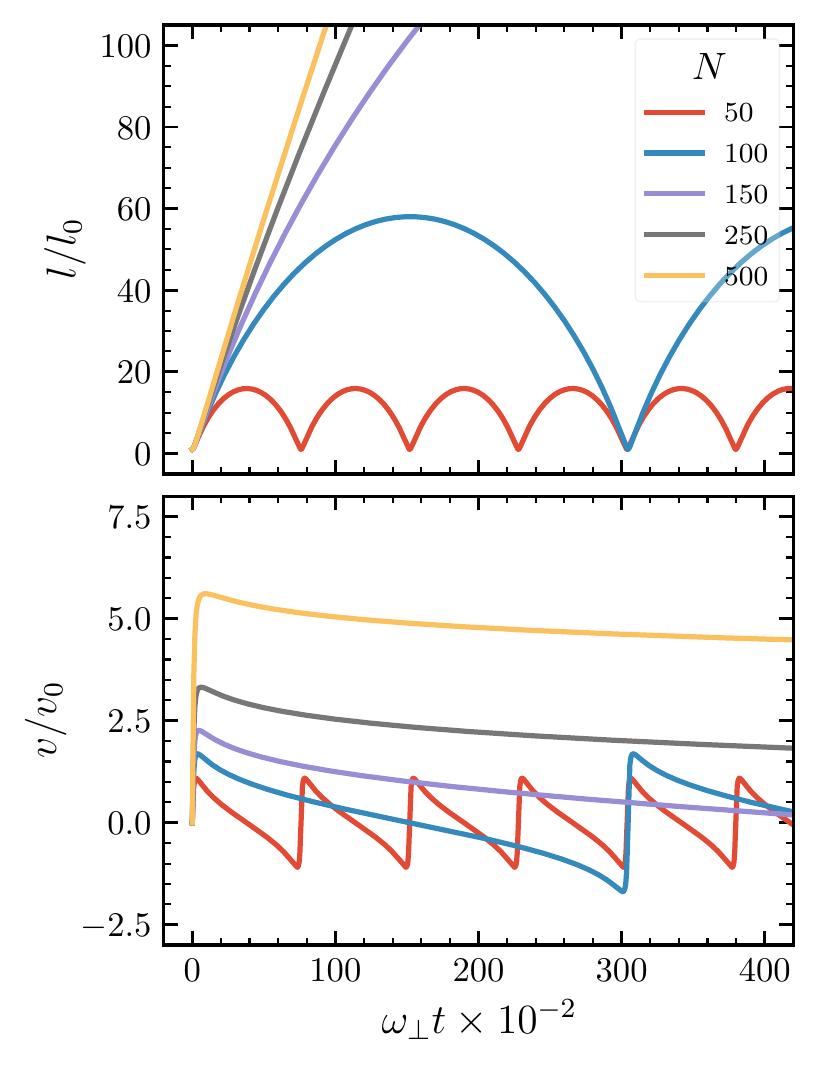}
  \caption{Size and velocity of the one--dimensional LogBEC in free expansion as a function of time
    $\omega_\bot t$ for several numbers of particles $N$. The scattering length is $\scatlen = 53.65
      a_0$, see Ref.~\cite{Gorlitz2001}, and $\beta = 2.2174 \times 10^{-3} (\hbar \omega_\bot)$.}%
  \label{fig:size-vel-sodium-01}
\end{figure}

\begin{figure}[t!]
  \centering
  \includegraphics[]{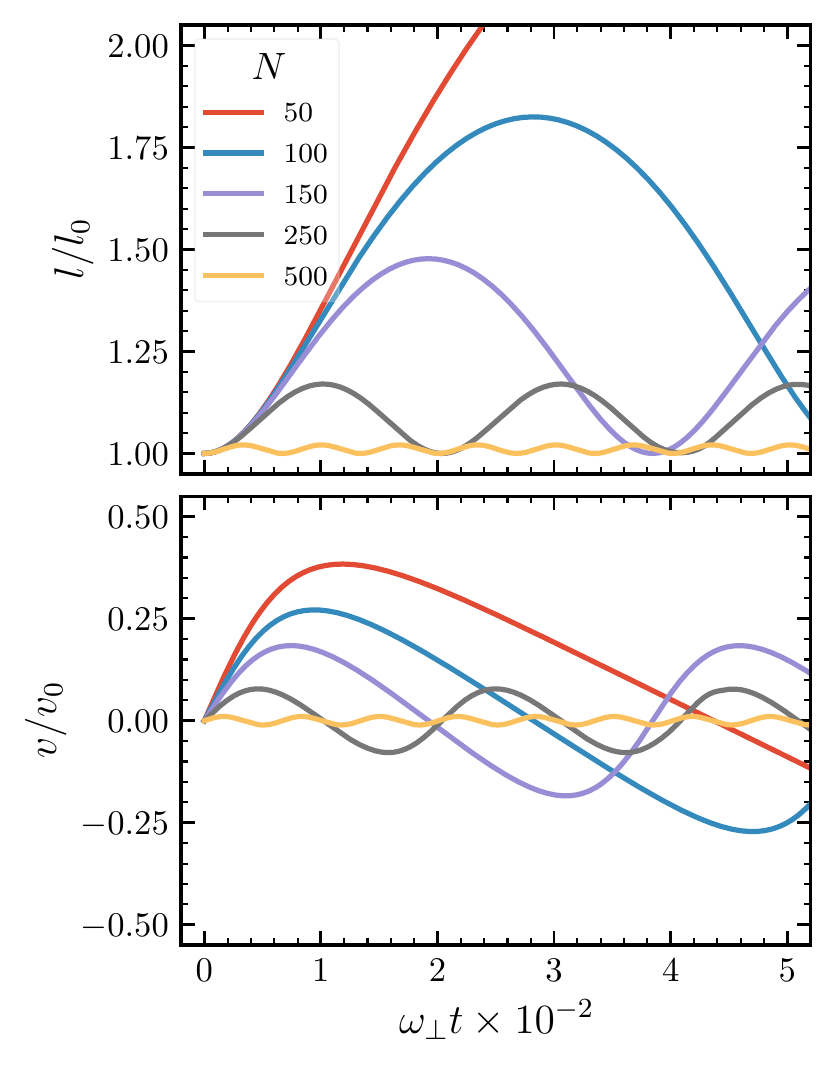}
  \caption{Size and velocity of the one--dimensional LogBEC in free expansion as a function of time
    $\omega_\bot t$ for several numbers of particles $N$. The scattering length is $\scatlen = -27.3
      a_0$, see Ref.~\cite{Bradley1995}, and $\beta = 2.2174 \times 10^{-3} (\hbar \omega_\bot)$.}%
  \label{fig:size-vel-lithium-01}
\end{figure}

\begin{figure}[t!]
  \centering
  \includegraphics[]{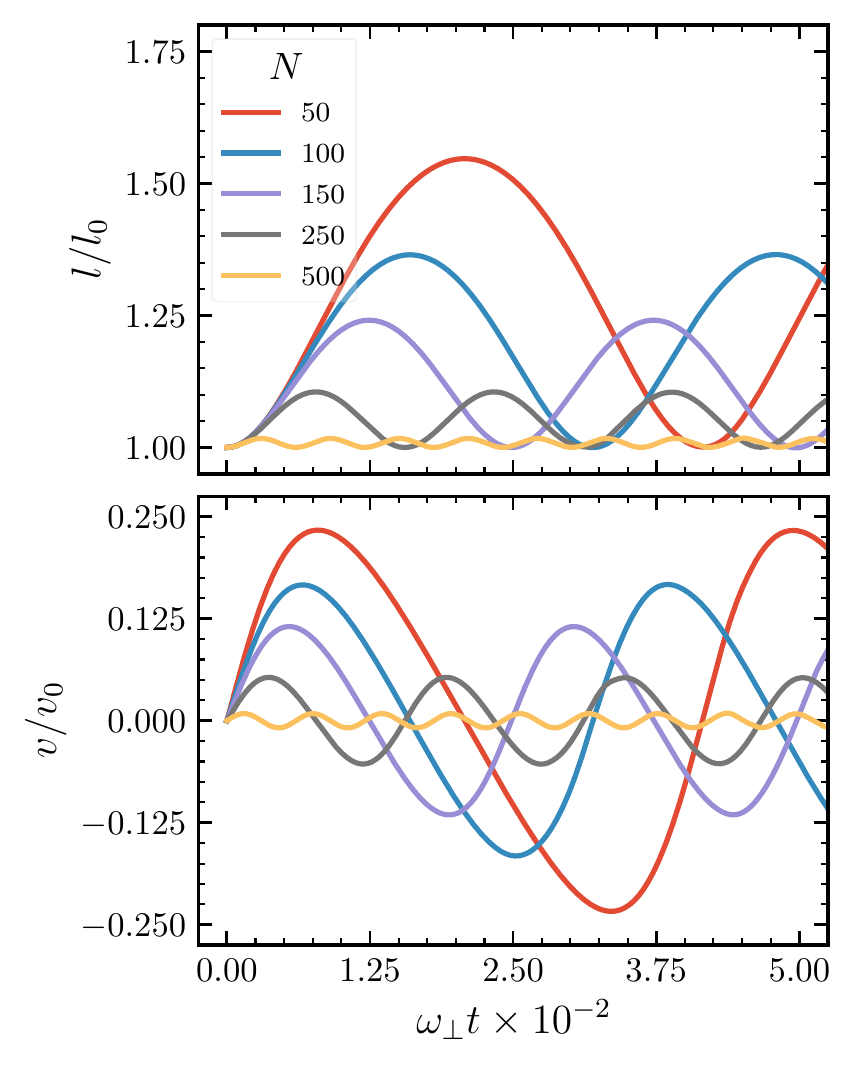}
  \caption{Size and velocity of the one--dimensional LogBEC in free expansion as a function of time
    $\omega_\bot t$ for several numbers of particles $N$. The scattering length is $\scatlen = -27.3
      a_0$, and $\beta = 4.4348 \times 10^{-3} (\hbar \omega_\bot)$.}%
  \label{fig:size-vel-lithium-02}
\end{figure}

\begin{figure}[t!]
  \centering
  \includegraphics[]{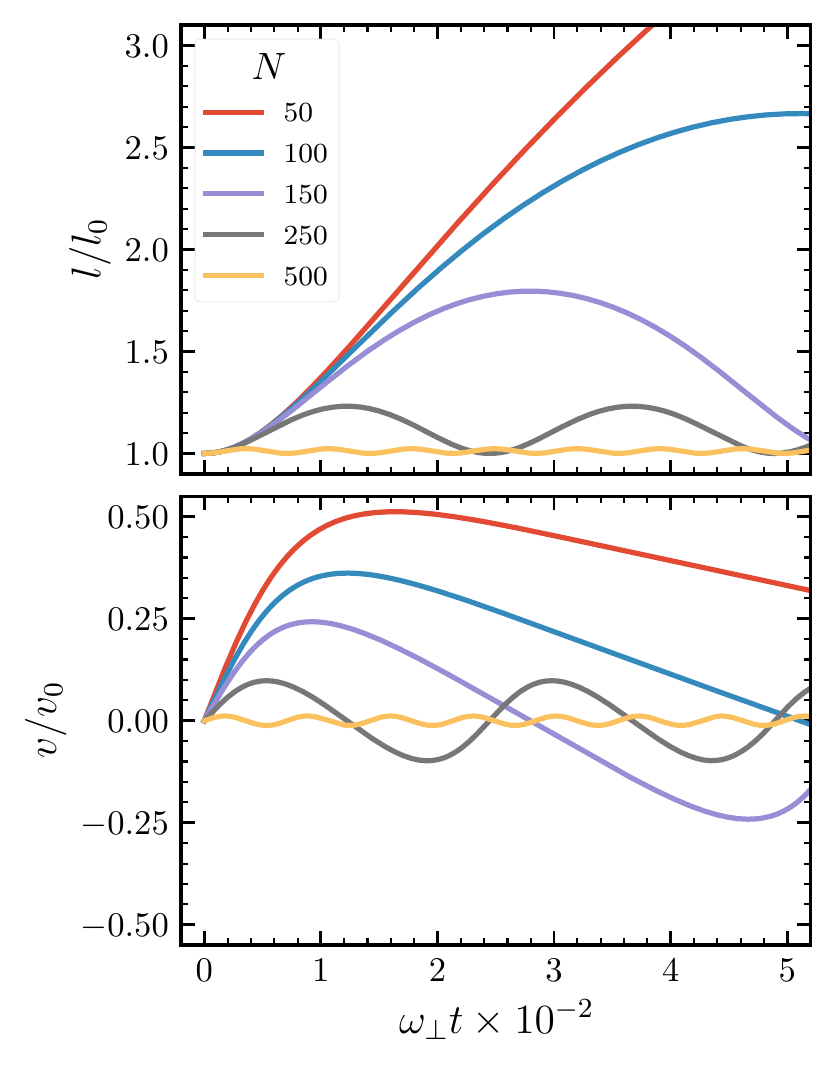}
  \caption{Size and velocity of the one--dimensional LogBEC in free expansion as a function of time
    $\omega_\bot t$ for several numbers of particles $N$. The scattering length is $\scatlen = -27.3
      a_0$, and $\beta = 1.1087 \times 10^{-3} (\hbar \omega_\bot)$.}%
  \label{fig:size-vel-lithium-03}
\end{figure}

\begin{figure}[t!]
  %\centering
  \includegraphics[]{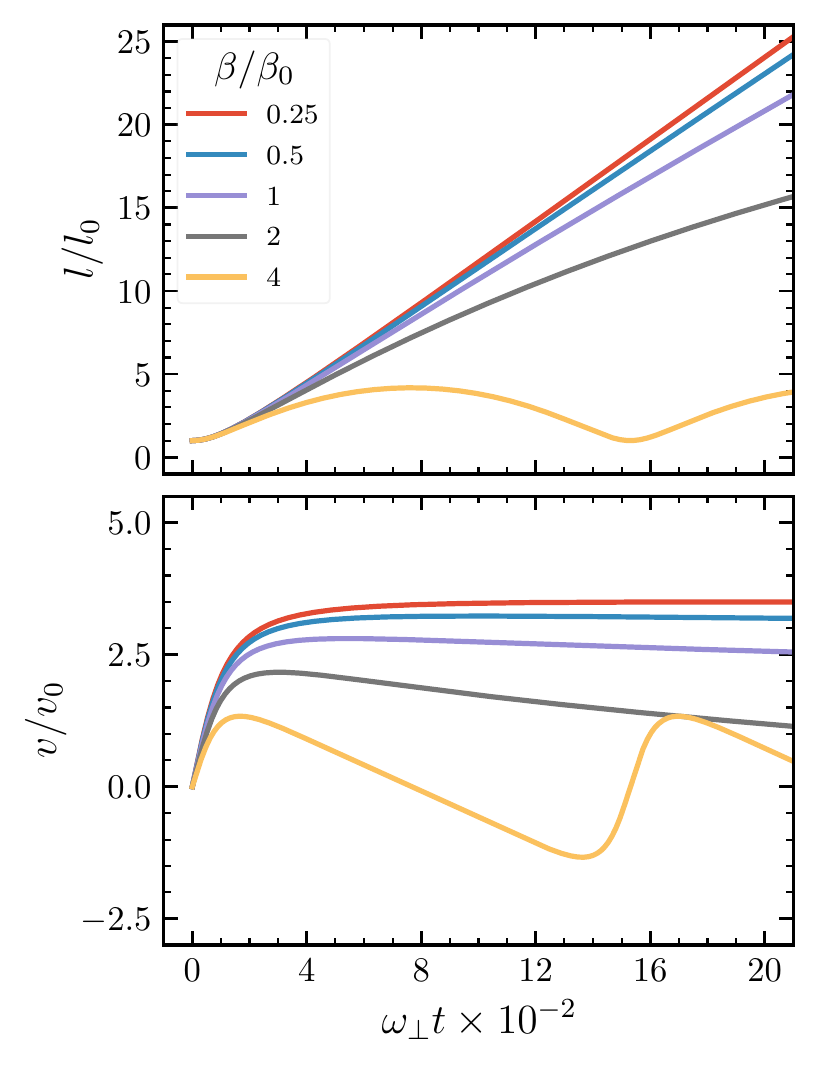}
  \caption{Size and velocity of the one--dimensional LogBEC in free expansion as a function of time
    $\omega_\bot t$ for different values of the nonlinear interaction $\beta$ and $N=200$. The
    scattering length is $\scatlen = 53.65 a_0$ and $\beta_0 = 2.2174 \times 10^{-3} (\hbar
      \omega_\bot)$.}%
  \label{fig:size-vel_sodium-beta-var_fig-011}
\end{figure}

\begin{figure}[t!]
  \centering
  \includegraphics[]{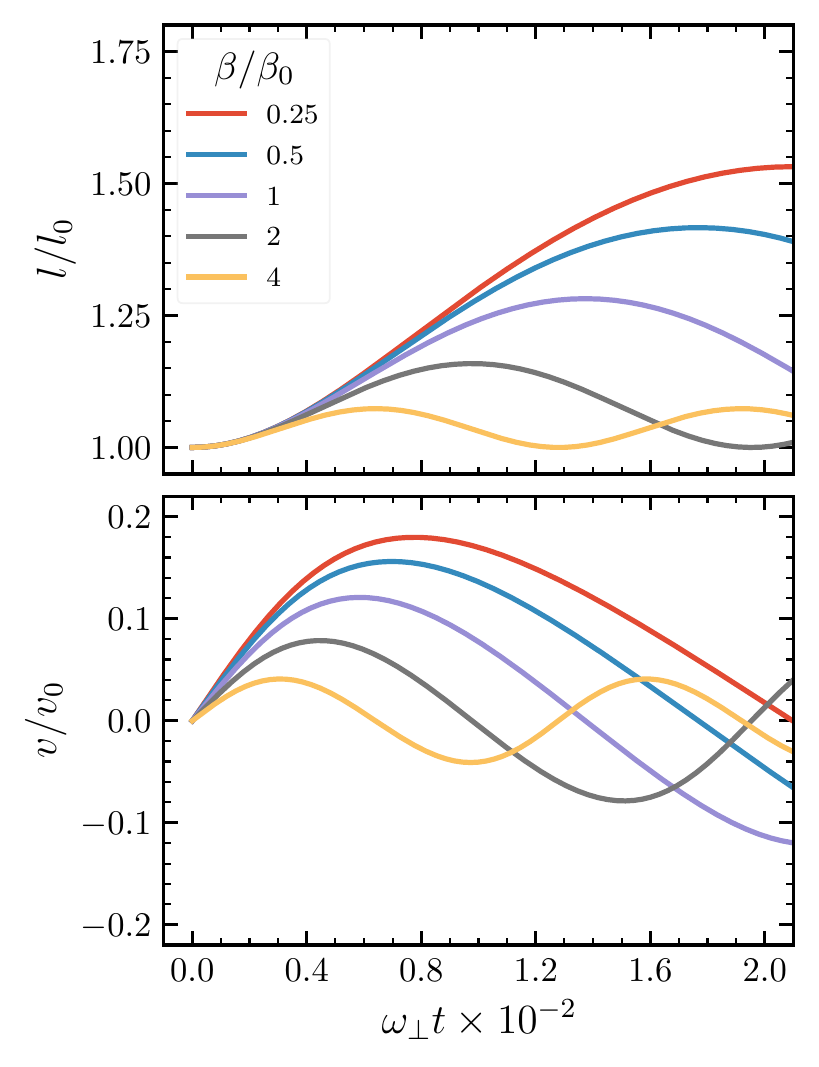}
  \caption{Size and velocity of the one--dimensional LogBEC in free expansion as a function of time
    $\omega_\bot t$ for different values of the nonlinear interaction $\beta$ and $N=200$. The
    scattering length is $\scatlen = -27.3 a_0$ and $\beta_0 = 2.2174 \times 10^{-3} (\hbar
      \omega_\bot)$.}%
  \label{fig:size-vel-lithium-012}
\end{figure}

After the external potential $V(z)$ is turned off, let us say, at $t=0$, there is a force that
changes the stability size of the cloud and produces an expansion of the {BEC}.
In order to determine an equation for the dynamics of the system, we must deduce the corresponding
kinetic energy $E_{F}$, Eq.~\eqref{eq:EFL}, as a function of time through its dependence on the radius
$l$ at any time.
Changing $l$ from its initial value to a new value $\tilde{l}$ amounts to a uniform dilation of the
system since the new density distribution $|\psi(z)|^{2}$ may be obtained from the old one
by changing the radial coordinate of each particle by a factor $\tilde{l}/l$, see, for instance,
Ref.~\cite{Pethick}.
Thus, the velocity of a particle can be expressed as follows,
\begin{equation}
  v(z)=z\,\frac{\dot{l}}{l}\,,
\end{equation}
where the dot means differentiation with respect to time.
Then, it is quite easy to obtain the kinetic energy $E_{F}$ by using the \textit{ansatz}
Eq.~\eqref{eq:TF}, with the result
\begin{equation}
  \label{eq:EF}
  E_{F}=\frac{m\,N}{8}\dot{l}^{2},
\end{equation}
which scales linearly with the number of particles $N$.
Moreover, assuming that the energy is conserved at any time, we obtain the following energy
conservation condition:
\begin{eqnarray}
  \nonumber
  && \frac{\dot{l}^{2}m}{8}+\frac{{\hbar}^{2}}{8m l^{2}}
  +\frac{g\,N}{4\sqrt{2\pi}\, l} -\frac{\beta}{4} \left(2 \ln \left(\frac{\alpha N}{l}\right)\right)\\
  \label{eq:EC} &=&
  \frac{{\hbar}^{2}}{8m l_{0}^{2}}+\frac{g\,N}{4\sqrt{2\pi}\, l_{0}}- \frac{\beta}{4} \left(2 \ln \left(\frac{\alpha N}{l_{0}}\right)\right),
\end{eqnarray}
where $l_{0}$ is the condensate's radius at time $t=0$, and $l$ corresponds to the radius at time
$t$.
Notice that in the ideal case, i.e., setting $g=\beta=0$, we obtain the analytical result
\begin{equation}
  \label{eq:usual}
  {l}^{2}={l}_{0}^{2}+{(v_{0}t)}^{2}.
\end{equation}
where $v_{0}={\hbar}/{m l_{0}}$ is the free velocity expansion of the condensate, corresponding to
the velocity predicted by Heisenberg's uncertainty principle for a particle confined a distance
$l_{0}$.
However, as we will see later in the manuscript, the contribution of parameters $g$ and $\beta$
modifies the free expansion properties.
When parameters $g$ and $\beta$ are present, the system dynamics allow the formation of quantum
droplets, which are not allowed in the ideal case or even when the contributions of $g$ alone are
taken into account for a single component.

To analyze the LogBEC's free expansion, we rewrite Eq.~\eqref{eq:EC} in terms of its rescaled size,
$r = l / a_\bot$, and its equilibrium size before turning off the trap, $r_0 = r(t=0) = l_0 /
  a_\bot$, as follows
\begin{align}
  \label{eq:bec-radius-rescaled}
  \nonumber {\left( \frac{d r}{d \tau} \right)}^2 & + \left( {r}^{-2} - {r_0}^{-2} \right) + \frac{4}{\sqrt{2\pi}} \left( N \frac{\scatlen}{a_\bot} \right) \left( r^{-1} - {r_0}^{-1}  \right) - \\
                                                  & - \frac{4\beta}{\hbar \omega_\bot} \ln \left(\frac{r}{r_0}\right) = 0,
\end{align}
where $\tau = \omega_\bot t$ is the rescaled time in terms of the transverse trapping frequency.
Notice that Eq.~\eqref{eq:bec-radius-rescaled} shows that the one--dimensional LogBEC radius does
not depend on $\alpha$.
By solving numerically Eq.~\eqref{eq:bec-radius-rescaled}, we can analyze the size of the
one--dimensional LogBEC under free expansion (keeping the parameter $\beta > 0$) for several numbers
of particles, considering that its physical characteristics are similar to those of $^{23}$Na and
$^{7}$Li.
The results of this analysis are shown in
Figs.~\ref{fig:size-vel-sodium-01},~\ref{fig:size-vel-lithium-01},~\ref{fig:size-vel-lithium-02}
and~\ref{fig:size-vel-lithium-03}.
Notice that, qualitatively speaking, the $^{23}$Na and the $^{7}$Li free--expansion behavior is very
similar, in the sense that they oscillate around a specific size that corresponds to the
equilibrium radius.
Moreover, the oscillation frequency of the cloud depends on the number of particles, i.e., the
oscillation increases for a larger number of particles.
Additionally, we observe that they form confined clouds in both cases, i.e., for $^{23}$Na and
$^{7}$Li, the so--called \textit{self–sustainability or self--confinement} appears.
In other words, the system is able to form oscillating quantum droplet--type configurations.
Specifically, we notice that in the $^{23}$Na case (see Fig.~\ref{fig:size-vel-sodium-01}), in which
$a_{S}>0$ and $\beta>0$, the time range expansion is significative larger, which is translated to
more slow oscillations.
Also, the cloud's size is much bigger compared with the case $a_{S}<0$ and $\beta>0$ for $^{7}$Li
showed in Figs.~\ref{fig:size-vel-lithium-01},~\ref{fig:size-vel-lithium-02}
and~\ref{fig:size-vel-lithium-03}.
In this scenario, we have taken two different values of $\beta$ compared to the upper value reported
in Ref.~\cite{Gahler1981}.
We can also see that a specific value for the parameter $\beta$ modifies the size and frequency of
the cloud oscillations, as shown in Figs.~\ref{fig:size-vel-lithium-02}
and~\ref{fig:size-vel-lithium-03}.
However, this choice for the parameter $\beta$ does not change the fact the that the system can form
quantum droplet--type configurations.
The quantum droplet--type configurations only disappear when $\beta$ vanishes, as expected.

Let us point out that the oscillation time range shown in the
Figs.~\ref{fig:size-vel-lithium-01},~\ref{fig:size-vel-lithium-02},
and~\ref{fig:size-vel-lithium-03} are around $100$ and $200$ milliseconds, which means that a
similar experiment could be performed on earth laboratory experiments~\cite{MR} where the free
expansion time is of the order of milliseconds.
In comparison, the oscillation time shown in Fig.~\ref{fig:size-vel-sodium-01} corresponds to a
physical time of approximately $3$ seconds.
Therefore, a similar experiment could be performed in a free fall microgravity environment like in
Refs.~\cite{MUT,MUT1}, or in a microgravity setup orbit experiments like in Ref.~\cite{NASA0,NASA}
if larger times in the free expansion of the cloud are necessary.

Finally, in Figs.~\ref{fig:size-vel_sodium-beta-var_fig-011} and~\ref{fig:size-vel-lithium-012}, we
show the system's behavior under free expansion for several values of the parameter $\beta$ in the
case of $^{23}$Na and $^{7}$Li with $N=200$ particles, correspondingly.
We immediately notice that both systems are stable for the chosen parameters.
Consequently, the \textit{self--sustainability} or \textit{self--confinement} (or specifically
speaking, the formation of oscillating quantum droplet--type configurations) also appears.
We have restricted our analysis to positive values of $\beta$ since negative values seem to have no
physical relevance, as was pointed out in Refs.~\cite{Gahler1981,IB}.
However, negative $\beta$ values could be analyzed in the present context since the only restriction
is the upper bound reported in Ref.~\cite{Gahler1981}.
Accordingly, a lower bound for $\beta$ could be estimated through free expansion experiments and its
corresponding quantum droplets if they appear.
Nevertheless, the study of negative values of $\beta$ deserves a more in--depth analysis and is out
of the present work scope.
% We only consider positive values , as it was mentioned above in the manuscript, negative values of $\beta$ deserve a more
% in--depth study to analyze the corresponding stability of the LogBEC and the eventual formation of
% quantum droplet--type configurations in this scenario.

\section{Conclusions}%
\label{sec:5}

In the present work, we prove that the one--dimensional LogBEC under free expansion can form quantum
droplet--type configurations.
In other words, we can prove that our model predicts almost the same structural configuration,
qualitatively speaking for the case of attractive and repulsive interactions encoded in the
one--dimensional two--body interaction parameter $g$, keeping $\beta>0$.
Additionally, we also prove with our formalism that it is not necessary a mixture of BECs, i.e., a
two--component system, to obtain \textit{self--sustaining or self--confined} configurations or, more
specifically, quantum droplets.
On the other hand, the results obtained here must be generalized, i.e., it is necessary to solve the
full LogGPE equation, within its regime of validity, to obtain a more general description of the
ground state wave function, its energy, and dynamics.
This kind of analysis would allow us to establish the validity of the \textit{ansatz} used in the
present work approach.
In this sense, it could be interesting to extend the formalism described in the present report to
explore, for instance, the possibility of negative $\beta$ values and how this affects the
free--expansion of the cloud together with its relation to the formation of quantum droplet-type
configurations
Also, the eventual appearance of emission of matter--wave jets, as reported in Ref.~\cite{Logan},
can be analyzed, in principle, without the requirement of Feshbach resonances in the case of dense
enough one--dimensional systems.
%and clearly, this leads to the possibility of extend the analysis reported here to
%three--dimensional LogBECs.
%
Following this line of thought, it also becomes quite exciting to study \textit{bosenova}--type
effects in three, two, and one-dimensional systems.
Finally,  it would be exciting to extend the study realized in the present work to explore phenomena
in the gravitational physics context.
For instance, we could study the stability of the so-called \textit{boson stars} and the formation
of matter jets in them.
We could also analyze the formation of eventual \textit{bosenovas} and explore the relation of
BECs as boson stars with dark matter in the universe~\footnote{See for instance Ref.~\cite{BS} and references
  therein for some insights related to this topic.}.
% Finally, from the gravitational physics point of view, it would be very interesting to explore the
% possibility to extend the study described in the present work in order to explore the stability of
% the so--called \textit{boson stars}, the formation of jets for these systems and the analysis of
% eventual \textit{bose novas}, and perhaps, include its relation with dark matter in the
% universe\footnote{See for instance Ref.~\cite{BS} and references therein for some insights related
% to this topic.}.
%\bigskip

\begin{acknowledgments}
  E. Castellanos acknowledges the receipt of the grant from the Abdus Salam International Centre for
  Theoretical Physics, Trieste, Italy. This work was partially supported also by CONACyT M\'exico
  under Grant No. 304001.
\end{acknowledgments}
%---------------------------------------------------------------------------------------------------------------------------------------------------------------------
%---------------------------------------------------------------------------------------------------------------------------------------------------------------------

\bibliographystyle{apsrev4-2}
\bibliography{references.bib}

% \begin{thebibliography}{99}

%  \bibitem{Ueda}
%  M. Ueda, Fundamentals and New Frontiers of Bose--Einstein Condensation
%  World Scientific, Singapore, 2010.
%
%
%  \bibitem{Pitaevski}
%  L. P. Pitaevskii and S. Stringari,  Bose-Einstein Condensation, Clarendon Press, Oxford
%  2003.

% \end{thebibliography}

%---------------------------------------------------------------------------------------------------------------------------------------------------------------------
%---------------------------------------------------------------------------------------------------------------------------------------------------------------------

\end{document}